\documentclass[usenatbib]{mn2e}
\usepackage{psfig}
\usepackage{amssymb}

\title[A giant molecular cloud falling through the heart of Cygnus
  A]{A giant molecular cloud falling through the heart of Cygnus A: 
  clues to the triggering of the activity}

\author[M.\,J.\ Bellamy et.\ al.] {M.\,J.\ Bellamy,$^1$\thanks{E-mail:
m.bellamy@sheffield.ac.uk} and C.\,N.\ Tadhunter.$^1$\\
$^1$Department of Physics and Astronomy, University of Sheffield,
Sheffield S3 7RH, UK}

\date{\center{\Large To be submitted for publication in the Monthly
Notices of the Royal Astronomical Society \\
\vspace{.5cm} \today}}

\begin{document}
\maketitle

\begin{abstract}
We present intermediate resolution near-IR long-slit spectroscopic
data for the nearby radio 
galaxy Cygnus A (3C 405) (obtained with the NIRSPEC
spectrograph on the Keck II telescope). The data reveal considerable
complexity in the near-IR emission line kinematics, including line
splittings of 200-350 km s$^{-1}$ and a mixture of narrow
(FWHM $\sim$200 km s$^{-1}$) and broad (FWHM $\sim$700 km s$^{-1}$)
components to the emission
lines. It is notable that the Pa$\alpha$ and H$_2$
emission lines show markedly different kinematics, both on- and
off-nucleus. Overall, the data provide evidence for the
presence of a giant molecular cloud falling through the heart of the
Cygnus A host galaxy, the motion of which is
not driven by the AGN itself. We suggest that this cloud may be
connected to the triggering of the activity in this highly powerful
AGN. We also detect split H$_2$ components on the nucleus that 
are likely to originate in the circum-nuclear torus.
\end{abstract}


\section{INTRODUCTION}
The prototype radio galaxy Cygnus A (3C 405) is the most luminous
extragalactic radio source in the local universe and the only nearby
(z=0.0558) radio galaxy of comparable power to 3C sources at
z$\sim$1. In consequence, it is one of
the most well-studied radio galaxies. Previous work has
concentrated on using Cygnus A to test the 
unified schemes for powerful radio galaxies
(\citealt{ueno94}; \citealt{ogle97}; \citealt{young02}) and
to search for signs of the impact of the activity on the host galaxy
in the form of AGN-driven outflows (\citealt*{taylor03}). Dynamical
studies have also led to 
estimates of the supermassive black hole mass (\citealt{tadhunter03}).


However, despite the large amount of progress that has been made over
the last decade in the understanding of Cygnus A and its host galaxy,
relatively little is known about the events that triggered the
activity. In this paper we present near-IR spectroscopic observations,
of higher resolution than previously published near-IR data
(\citealt{ward91}; \citealt*{tsr99}; \citealt{wilman00}),
that provide evidence for non-equilibrium gas motions that may have a
direct bearing on the triggering of the activity in this poweful AGN.

In order to put our observations in context, we start by summarising
our current state of knowledge of the geometry and kinematics of the
near-nuclear regions of Cygnus A.

\subsection{Geometry and inclination of the near-nuclear regions}
HST imaging has revealed an
edge-brightened biconical structure on the
scale of a few hundred parsecs (\citealt{jackson98};
\citealt{tadhunter99}). This structure is most likely a direct result
of the nuclear-driven winds hollowing out `ionisation cones' in the
kpc-scale dust lane in which the AGN is embedded.
Moreover, the images also show evidence for a starburst ring
associated with the dust lane (\citealt{fosbury99}).
From these data, the best estimates for the
geometry of the biconical structure are opening half-angle
$\theta_{\frac{1}{2}}$ 
$\simeq$ 60\degr and inclination of the cone axis relative to the
line-of-sight i $\simeq$ 30\degr with the NW cone oriented
towards us and the SE cone oriented away from us
(see \citealt{tadhunter03} for a full discussion). From this geometry
we know that we 
are likely to be viewing Cygnus A very close to the opening angle of
the NW cone, but we also 
know that i $<$ (90 - $\theta_{\frac{1}{2}}$) as we do not have a
direct view of the quasar nucleus at optical wavelengths.


\subsection{Kinematic components in the near-nuclear regions}
The central regions of Cygnus A show considerable complexity in their
emission line kinematics. The various kinematic components 
can be tied to different processes associated with the AGN as follows:

\subsection*{Gravitational motions}

{\bf Kpc-scale disc:} near-IR Keck II spectroscopic data shows
  evidence for a rotating disc associated with the 3 kpc scale dust
  lane. The gas is in Keplerian rotation about a stellar core and an
  unresolved point mass of (2.5 $\pm$ 0.7) $\times$ 10$^9$ solar
  masses. The rotation axis of the gas is aligned at $\sim$9\degr to
  that of the large-scale radio axis (\citealt{tadhunter03}).

\vspace{2mm}
\noindent{\bf 300pc-radius disc:} Higher spatial
  resolution HST/STIS 
  optical data of \citet{tadhunter03} show evidence for a smaller disc
  structure. The lines from this disc are significantly
  broadened throughout (FWHM 500-900 km s$^{-1}$). Although the slit positions
  closest to the nucleus show evidence for rotation about a central
  black hole with mass similar to that deduced from the KeckII/NIRSPEC
  data, there is evidence for deviations from pure circular motions in
  the NW cone.

\subsection*{Evidence for outflows}

{\bf Emission line outflows:} Intermediate spectral resolution
  optical data of the [O III]$\lambda$5007 line (\citealt{taylor03})
  provide evidence for outflowing gas in the ionisation cones. A
  $\sim$300 km s$^{-1}$ outflow is detected in the NW cone
  (blueshifted), whereas a $\sim$400 km s$^{-1}$ outflow is detected
  in the SE cone (redshifted). These results are consisitent with gas
  being driven out of the cones by AGN-induced winds.

\vspace{2mm}
\noindent{\bf Extreme outflow:} an extreme outflow component is detected
  in [O III]$\lambda\lambda$(5007,4949) emission in the NW cone
  (\citealt{tadhunter91}). This component lies on the radio axis
  itself and is blueshifted by 1300 - 1800 km s$^{-1}$; it is likely
  to represent gas entrained in the outflowing radio plasma.

\vspace{2mm}
\noindent{\bf Scattering outflow:} The redshifted [O III]$\lambda$5007
  feature detected in polarised light, in both cones and the nucleus, by
  \citet{vanbemmel03} is consistent with the presence of a scattering
  outflow with velocity in the range 150-450 km s$^{-1}$.

\subsection*{Evidence for inflows}

{\bf Infalling H I:} Two components of H I 21cm absorption are detected
  in the nuclear regions of Cygnus A (\citealt{conway95}). In the rest
  frame, defined by the narrow component of Pa$\alpha$ in the nuclear
  aperture (see below), these components have {\it redshifts} of 227
  $\pm$ 9 km s$^{-1}$ and 48 $\pm$ 9 km s$^{-1}$. Since these
  absorption components are detected against the radio core, and
  are therefore in the foreground, they must be associated with
  infalling material.

\vspace{2.5mm}
\noindent Although there is now substantial evidence for the
gravitational motions and AGN-induced outflows in Cygnus A, the
evidence for inflowing material is relatively sparse and rests solely
in the H I observations of \citet{conway95}. However, it is the inflow
component that is most likely to provide the strongest clues to the
triggering of the activity. Therefore it is important to obtain
further information about the kinematics and distribution of the
inflowing gas. In this context we note that \citet{canalizo03} have
recently reported the discovery of a red, secondary
point source in the nuclear regions, $\sim$400pc southwest
of the radio nucleus. They argue that this secondary nucleus
represents the debris of a minor merger that
may be fuelling the AGN activity.

We assume the cosmological parameters of H$_0$=75 km s$^{-1}$
Mpc$^{-1}$ and q$_0$=0 throughout this paper. For these
parameters 1.00 arcsec corresponds to 1.00 kpc at the redshift of
Cygnus A.

\section{DATA COLLECTION AND REDUCTION}
The K-band near-IR spectra of Cygnus A were taken on 2000 May 22 using
the NIRSPEC spectrograph in grating mode on the Keck II telescope at
the Mauna Kea complex. Long-slit spectra were obtained with the slit
oriented along the radio axis (PA105)
and along PA180: in each case the slit was centred on the
nucleus. Velocity curves derived from single-Gaussian fits to these
data - which provide evidence for a rotating kpc-scale disc - are
presented in \citet{tadhunter03}. In this paper we will be
concerned only with the radio axis position angle (PA105). Four
exposures of 400 seconds were
obtained (at each slit position) with the exposures nodded in an ABBA
pattern to facilitate sky subtraction.

The spectra were rectified using purpose-written \textsc{iraf}
routines, and wavelength-calibrated in \textsc{iraf} using an argon/neon arc
lamp exposure taken at the time of the observations. Following the
manual removal of cosmic rays using the \textsc{clean} routine in
\textsc{figaro}, the separate exposures for each PA were combined
to produce a sky-subtracted two-dimensional spectrum. Telluric
features were removed by dividing a high S/N spectrum of the A0 standard
star HD203856, which was observed at the same airmass, into the galaxy
spectra. Use of the 0.57-arcsec slit resulted in a spectral
resolution of 10.9 $\pm$ 0.5 \AA~(140-170 km s$^{-1}$ for these data),
with the observations covering the wavelength range
19100-23000\AA. From measurements of stars along 
the slit, we estimate a plate scale
of 0.178 $\pm$ 0.004 arcsec pixel$^{-1}$, and an effective seeing of
0.75 $\pm$ 0.05 arcsec (FWHM) for the observations. Measurements of
night-sky emission lines in the reduced frames demonstrate that the
wavelength scale is accurate to within $\pm$ 0.5 \AA~($\pm$ 8 km
s$^{-1}$) along the full length of the slit.

The exposures of HD203856 were used to flux-calibrate the data, with
the assumption that the intrinsic spectral energy distribution (SED)
of the star is that of a perfect black-body at T = 9480 K. The
magnitude-to-flux conversion was performed with reference to
\citet*{bessell98}. The Starlink \textsc{figaro} and \textsc{dipso}
packages were used to analyse/reduce the data and fit the spectral
lines.

\section{RESULTS}

\begin{table*}
\begin{center}\begin{tabular}{llllll}
\hline
$\lambda$ Observed (\AA)&$\lambda$ Rest (\AA)&Species&FWHM (km
s$^{-1}$)&z&Flux (erg s$^{-1}$ cm$^{-2}$) \\
\hline
19382.5 $\pm$ 0.5&18358&H$_2$ $\nu$=1-0
S(5)&420 $\pm$ 20&0.05586 $\pm$ 0.00003&(8.0 $\pm$ 0.3)$\times$10$^{-16}$ \\
$\sim$19684&18639&He II (6-5)&$\sim$420&assumed 0.056&(4.61 $\pm$ 0.24)$\times$10$^{-16}$ \\
$\sim$19737&18691&He I (4F-3D, 3Fo-3D)&$\sim$420&assumed 0.056&(5.18 $\pm$ 0.25)$\times$10$^{-16}$ \\
19802.3 $\pm$ 0.5&18756.13&Pa$\alpha$ (Narrow)&250 $\pm$ 6&0.05583
$\pm$ 0.00003&(2.54 $\pm$ 0.13)$\times$10$^{-15}$ \\
19802.7 $\pm$ 0.5&18756.13&Pa$\alpha$ (Broad)&750 $\pm$ 14&0.05585
$\pm$ 0.00003&(5.62 $\pm$ 0.12)$\times$10$^{-15}$ \\
19947.4 $\pm$ 2.4&18920&H$_2$ $\nu$=1-0 S(4) (Component 1)$^a$&(270 $\pm$
70)$^{a}$&0.0544 $\pm$ 0.0001&(4.5 $\pm$ 2.1)$\times$10$^{-17}$ \\
19970.2 $\pm$ 2.4&18920&H$_2$ $\nu$=1-0 S(4) (Component 2)$^a$&(160 $\pm$
20)$^a$&0.0556 $\pm$0.0001&(9.8 $\pm$ 4.5)$\times$10$^{-17}$ \\
19979.1 $\pm$ 2.4&18920&H$_2$ $\nu$=1-0 S(4) (Component 3)$^a$&(370 $\pm$
40)$^a$&0.0562 $\pm$ 0.0001&(2.3 $\pm$ 0.5)$\times$10$^{-16}$ \\
20283 $\pm$ 1&19200$^{c,d}$&[S XI] (Narrow)&220 $\pm$ 50&$\sim$0.056$^d$&(3.0 $\pm$ 1.7)$\times$10$^{-16}$ \\
20287 $\pm$ 3&19200$^{c,d}$&[S XI] (Broad)&525 $\pm$ 100&$\sim$0.057$^d$&(3.9 $\pm$ 1.7)$\times$10$^{-16}$ \\
20528 $\pm$ 3&19451&Br$\delta$ (Broad)&560 $\pm$ 80&0.0555 $\pm$
0.0002&(2.7 $\pm$ 0.6)$\times$10$^{-16}$ \\
20537 $\pm$ 1&19451&Br$\delta$ (Narrow)&200 $\pm$ 30&0.05586 $\pm$
0.00005&(1.8 $\pm$ 0.6)$\times$10$^{-16}$ \\
20642 $\pm$ 1&19576&H$_2$ $\nu$=1-0 S(3) (Component 1)$^a$&(270 $\pm$
70)$^a$&0.05450 $\pm$ 0.00005&(1.6 $\pm$ 0.5)$\times$10$^{-16}$ \\
20665 $\pm$ 1&19576&H$_2$ $\nu$=1-0 S(3) (Component 2)$^a$&(160 $\pm$
20)$^a$&0.05568 $\pm$ 0.00005&(4.6 $\pm$ 2.0)$\times$10$^{-16}$ \\
20674 $\pm$ 1&19576&H$_2$ $\nu$=1-0 S(3) (Component 3)$^a$&(370 $\pm$
40)$^a$&0.05629 $\pm$ 0.00005&(8.9 $\pm$ 2.3)$\times$10$^{-16}$ \\
20728 $\pm$ 1&19634&[Si VI] (Narrow)$^b$&(250 $\pm$ 6)$^b$&0.05577 $\pm$
0.00005&(6.3 $\pm$ 0.3)$\times$10$^{-16}$ \\
20728 $\pm$ 1&19634&[Si VI] (Broad)$^b$&(750 $\pm$ 14)$^b$&0.05577 $\pm$
0.00005&(1.39 $\pm$ 0.03)$\times$10$^{-15}$ \\
21445 $\pm$ 1&20338&H$_2$ $\nu$=1-0 S(2) (Component 1)$^a$&(270 $\pm$
70)$^a$&0.05448 $\pm$ 0.00005&(5.8 $\pm$ 1.9)$\times$10$^{-17}$ \\
21470 $\pm$ 1&20338&H$_2$ $\nu$=1-0 S(2) (Component 2)$^a$&(160 $\pm$
20)$^a$&0.05575 $\pm$ 0.00005&(1.7 $\pm$ 0.7)$\times$10$^{-16}$ \\
21480 $\pm$ 1&20338&H$_2$ $\nu$=1-0 S(2) (Component 3)$^a$&(370 $\pm$
40)$^a$&0.05630 $\pm$ 0.00005&(3.2 $\pm$ 0.8)$\times$10$^{-16}$ \\
21580 $\pm$ 3&20412$^c$&H$_2$ $\nu$=8-6 O(3)&660 $\pm$ 90&0.0573 $\pm$
0.0002&(1.7 $\pm$ 0.2)$\times$10$^{-16}$ \\
21736 $\pm$ 1&20587&He I (2P-2S, 1Po-1S)&445 $\pm$ 30&0.05586 $\pm$
0.00005&(3.2 $\pm$ 0.2)$\times$10$^{-16}$ \\
21895 $\pm$ 2&20735&H$_2$ $\nu$=2-1 S(3)&445 $\pm$ 50&0.0560
$\pm$ 0.0001&(1.7 $\pm$ 0.2)$\times$10$^{-16}$ \\
22374 $\pm$ 3&21218&H$_2$ $\nu$=1-0 S(1) (Component 1)&270 $\pm$
70&0.0546 $\pm$ 0.0001&(1.5 $\pm$ 0.5)$\times$10$^{-16}$ \\
22401 $\pm$ 1&21218&H$_2$ $\nu$=1-0 S(1) (Component 2)&160 $\pm$
20&0.05579 $\pm$ 0.00005&(4.0 $\pm$ 1.7)$\times$10$^{-16}$ \\
22410 $\pm$ 3&21218&H$_2$ $\nu$=1-0 S(1) (Component 3)&370 $\pm$
40&0.0563 $\pm$ 0.0001&(7.8 $\pm$ 2.0)$\times$10$^{-16}$ \\
22865 $\pm$ 2&21661&Br$\gamma$ (Narrow)&350 $\pm$ 70&0.0557 $\pm$
0.0001&(3.5 $\pm$ 1.8)$\times$10$^{-16}$ \\
22872 $\pm$ 6&21661&Br$\gamma$ (Broad)&700 $\pm$ 130&0.0560 $\pm$
0.0003&(3.6 $\pm$ 1.9)$\times$10$^{-16}$ \\
\hline

\end{tabular}\end{center}

\caption{Results of the \textsc{dipso} fits to the spectral lines in
  the PA105 central aperture. The FWHMs have been corrected for the
  instrumental width. The redshifts have been heliocentrically
  corrected.\newline${^a}$Fit from the H$_2$ $\nu$=1-0 S(1)
  components.\newline${^b}$Fit from the Pa$\alpha$
  components.\newline${^c}$Rest frame wavelengths from
  \citet{tsr99}. \newline${^d}$Rest wavelength not known accurately
  from atomic physics.}
\label{table: results}
\end{table*}

\begin{figure}

\centerline{\psfig{figure=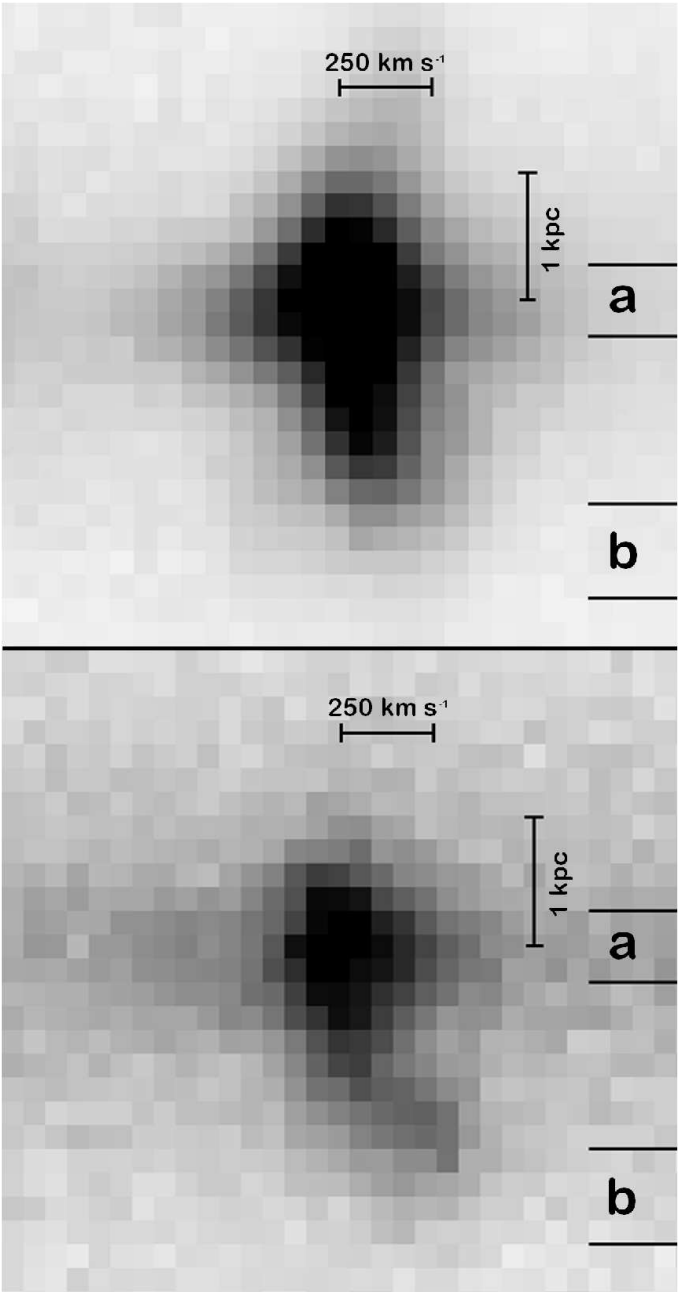,width=9cm,angle=0.}}

\caption{The 2D profiles of the Pa$\alpha$ (top) and H$_2$ $\nu$=1-0
  S(1) (bottom) emission in the PA105 data. The spatial scale runs
  vertically with the NW side at the bottom; the velocity scale runs
  horizontally with positive velocities to the right. The extraction
  apertures are marked on the right hand side: a) the nuclear
  aperture and b) the extended NW aperture.} 

\label{fig: h2}

\end{figure}

\begin{figure}

\centerline{\psfig{figure=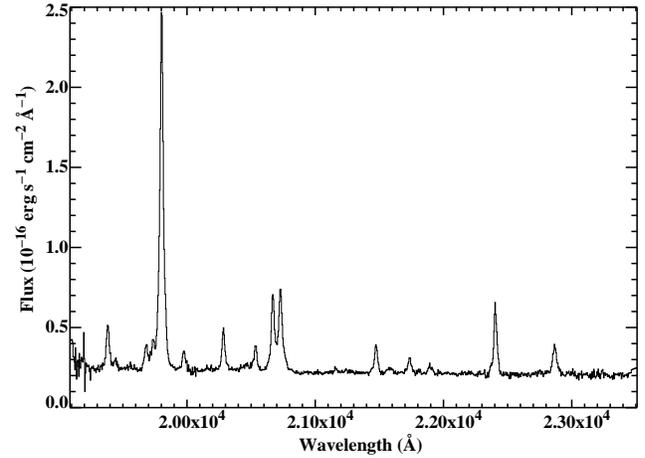,width=9cm,angle=0.}}

\caption{K-band 1D spectrum extracted from the central aperture of the
PA105 data.}

\label{fig: data}

\end{figure}

To examine the kinematics along the slit with good S/N,
a number of apertures were extracted. Two of these apertures reveal
the complex kinematics that are discussed in this paper. The first
aperture is centred on the nucleus (to the nearest pixel) and has a
width of three pixels ($\sim$540 pc). The second aperture is centred
$\sim$1~880 pc to the NW of the nucleus and is four pixels ($\sim$720
pc) in width. Fig. \ref{fig: h2} shows the greyscale 2D profiles of
the Pa$\alpha$ and the H$_2$ $\nu$=1-0 S(1) line emission with these
apertures overlaid. Fig. \ref{fig: data} shows the 1D spectrum
extracted from the PA105 central aperture.

The emission lines in the aperture spectra were fitted with Gaussian
profiles using the \textsc{dipso} package. Each line was fitted with
the minimum number of Gaussians required to give an acceptable
fit. Table \ref{table: results} shows the results of the line
fitting. Several lines show evidence for 
multiple kinematic components (notably the brighter H$_2$ lines) and
these components have been fitted separately. We see no evidence for the
H$_2$ $\nu$=2-1 S(2) line seen by \citet{wilman00}, but note that
their slit is approximately twice the width of the slit used in the
acquisition of these Keck II data. This may suggest that the H$_2$
$\nu$=2-1 S(2) emission originates in a more extended region.

The Pa$\alpha$ and molecular hydrogen lines are discussed in more
detail in the following. As previously noted by \citet{wilman00}, the
Pa$\alpha$ and molecular hydrogen lines exhibit different properties
and are therefore discussed in separate sections.
Where velocity scales and shifts are given, the zero velocity
corresponds to the heliocentric z=0.05583, the centre of the
Pa$\alpha$ narrow component in the nuclear aperture. This systemic
redshift is consistent with, though slightly lower than that deduced
from the Pa$\alpha$/H$_2$ rotation curves (\citealt{tadhunter03}:
z=0.05592 $\pm$ 0.00005). 

\subsection{Paschen alpha}
The Pa$\alpha$ emission exhibits kinematics, both on- and
off-nucleus, that show similarities to those of the optical [O III]
lines (\citealt{taylor03}). Unlike the case of the
molecular hydrogen lines (see section 3.2), there are no comparable
recombination lines with which to 
check the accuracy of the component fittings. However, the Pa$\alpha$
line is the brightest line in the spectrum, and has a high S/N
ratio, so we can be confident in the results of the fittings.

In the central nuclear aperture,
the Pa$\alpha$ line clearly has broad wings that cannot be fitted with
a single Gaussian profile. Fitting shows a central narrow component
with FWHM = 250 km s$^{-1}$ and a much broader component with FWHM =
750 km s$^{-1}$. HST/STIS data of the nuclear regions shows 
a rotation-curve in the [O III] line. Integrated over the inner 600 pc, and
fitted with a single-Gaussian, the profile of the [O III] line gives a
FWHM of 850 $\pm$ 50 km s$^{-1}$(\citealt{tadhunter03}). The broad
Pa$\alpha$ component is therefore likely to be due to a combination of
unresolved rotation and turbulent motion in the $<$300 pc-scale
nuclear disc. The centre of the broad component is coincident
with that of the narrow component within 1$\sigma$ 
uncertainty.

\begin{figure}

\centerline{\psfig{figure=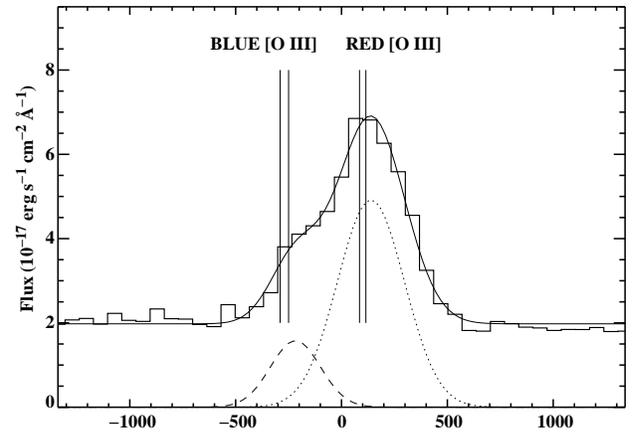,width=9cm,angle=0.}}
\centerline{\psfig{figure=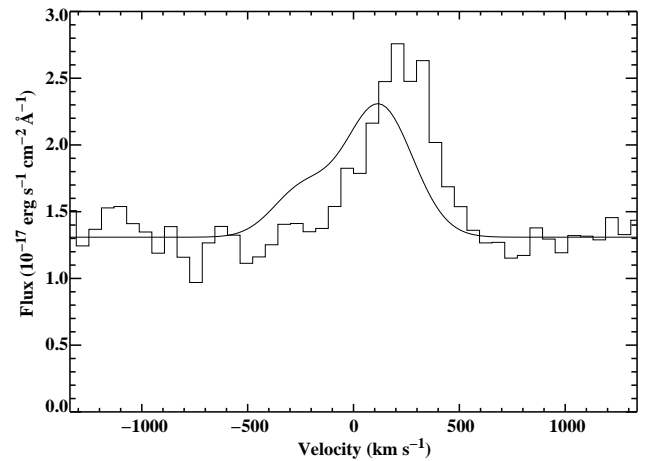,width=9cm,angle=0.}}

\caption{(Top) The two components of Pa$\alpha$ in the extended NW
  aperture (1 880 pc from the nucleus). Marked
  on the graph are the velocity shifts of the [O III] components in
  the same region (\citealt{taylor03}). (Bottom) The two
  Pa $\alpha$ components  
  shown fitted to the H$_2$ $\nu$=1-0 S(1) line in the NW aperture. The
  relative fluxes of the Pa$\alpha$ fit are retained as the fitting
  could not be made with these parameters free.}
\label{fig: figpaaex}

\end{figure}

The situation is different in the extended NW aperture in the
PA105 data: this aperture is centred 1 880 pc NW of the
nucleus, along the radio axis in the NW ionisation cone. In this
aperture the Pa$\alpha$ is split into two narrow components (FWHM $\sim$360
km s$^{-1}$), one of which is significantly blueshifted by 240 $\pm$ 30
km s$^{-1}$ in the galaxy-frame (see Fig. \ref{fig: figpaaex}); the other,
brighter, component is redshifted by 120 $\pm$ 13 km s$^{-1}$. As can
be seen in
Fig. \ref{fig: figpaaex}, these components provide a poor fit to
the H$_2$ emission in the same aperture. In the same spatial region,
two components are 
seen in the [OIII] $\lambda$5007 line at optical wavelengths: these
are shifted by (-270 $\pm$ 20) and (+100 $\pm$ 15) km s$^{-1}$ respectively
(\citealt{taylor03}). Allowing for the different slit widths and
systematic uncertainties, the Pa$\alpha$ components are entirely
consistent with the [O III] components of \citet{taylor03}. Therefore
the blueshifted component is likely to be associated with the emission
line outflow in the NW cone.

\subsection{Molecular hydrogen}

\begin{figure}

\centerline{\psfig{figure=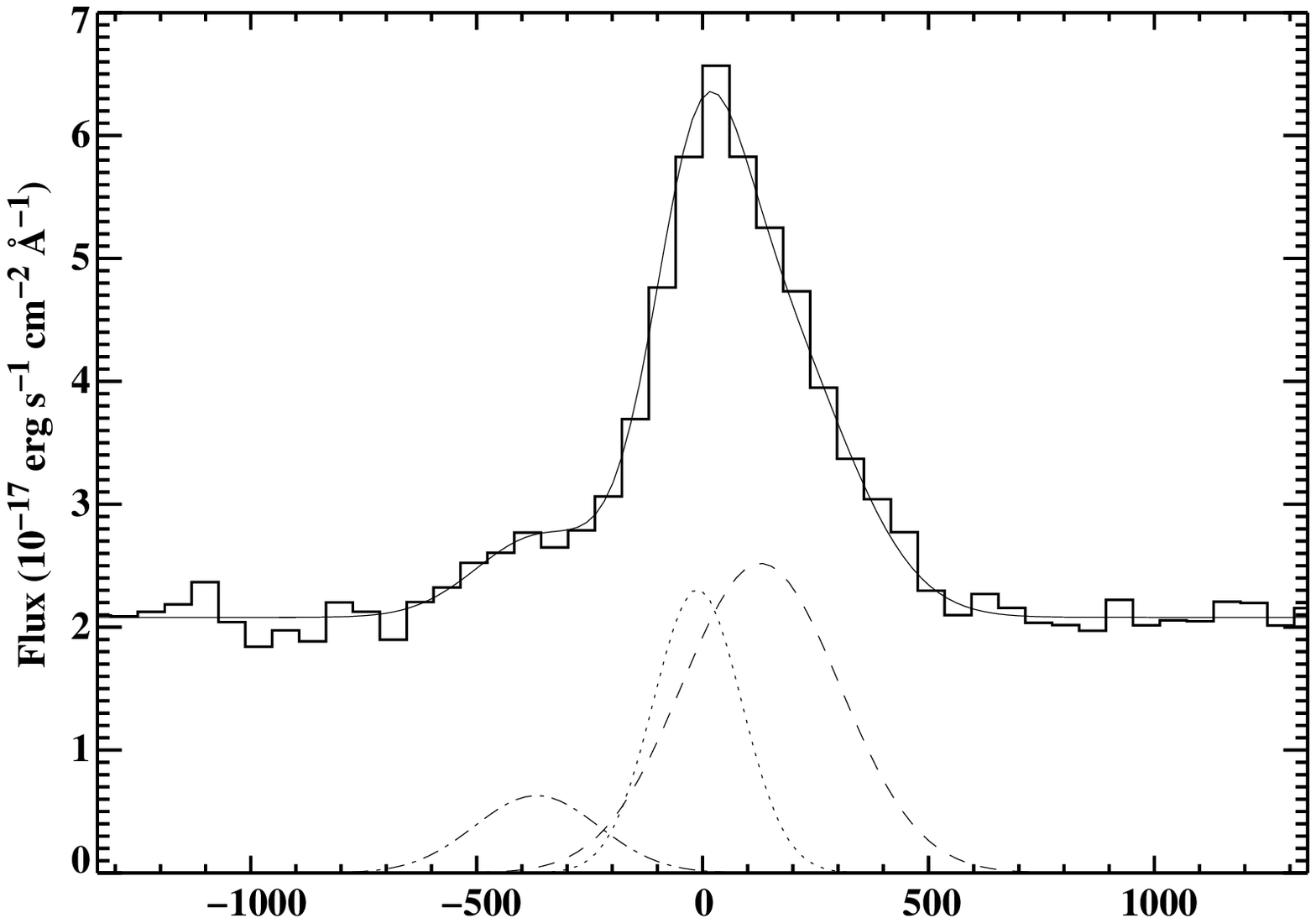,width=9cm,angle=0.}}
\centerline{\psfig{figure=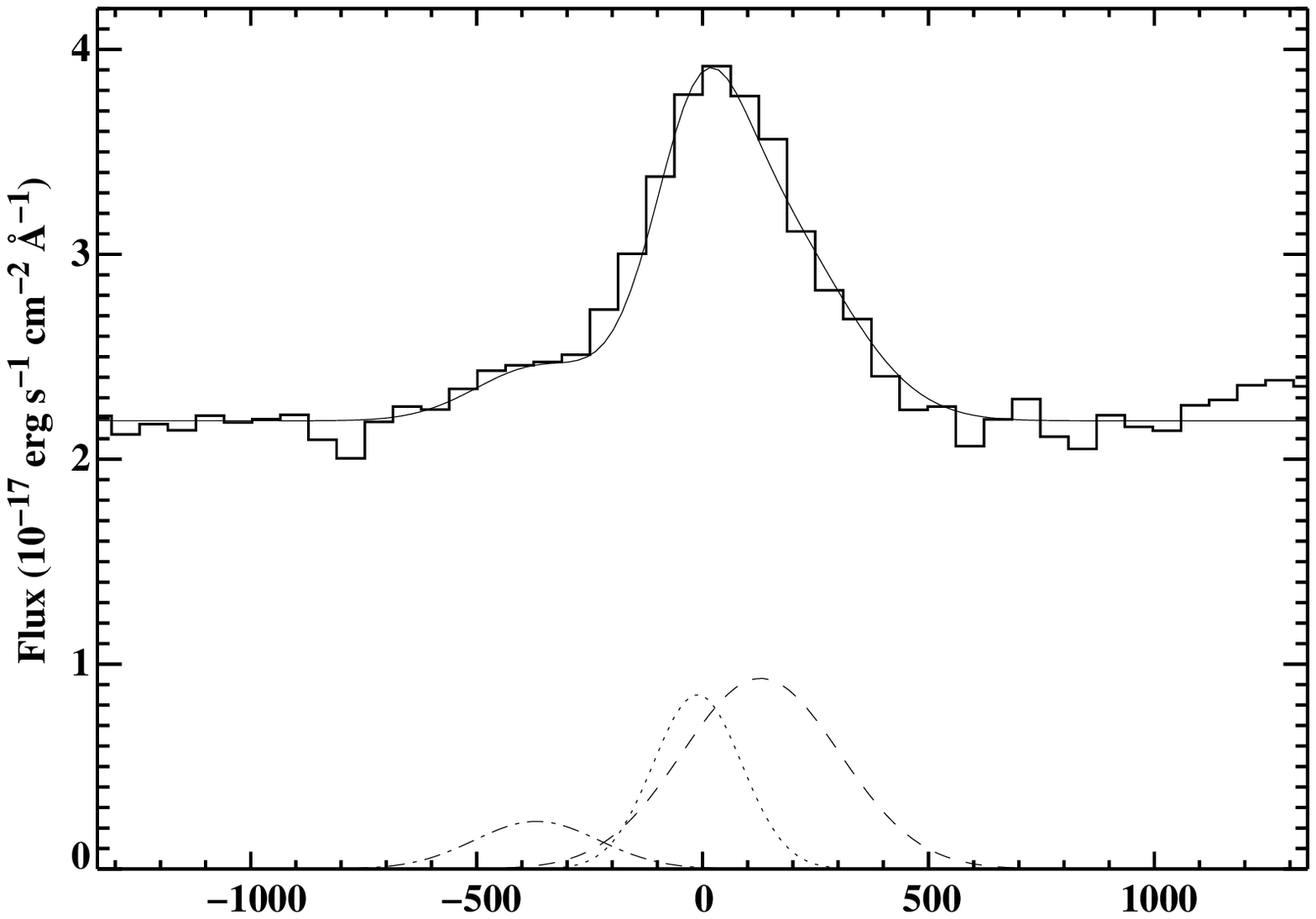,width=9cm,angle=0.}}
\centerline{\psfig{figure=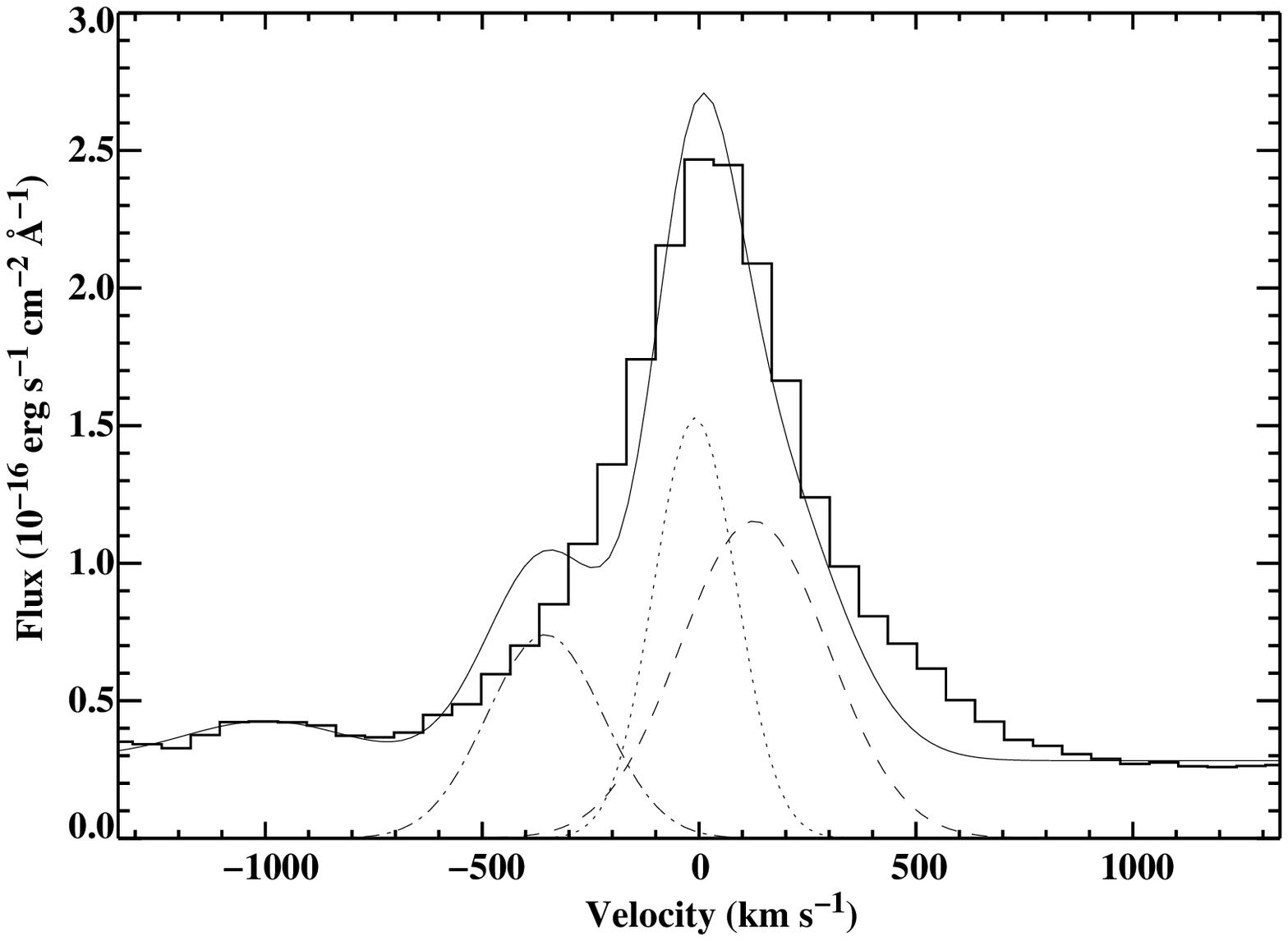,width=9cm,angle=0.}}

\caption{The three components (see text) of the H$_2$ lines as shown
  fitted to (top) the H$_2$ $\nu$=1-0 S(1) line, (middle) the H$_2$ $\nu$=1-0
  S(2) line and (bottom) the Pa$\alpha$ line (all in the nuclear
  aperture). The weak feature redward of the H$_2$ $\nu$=1-0 S(2)
  line is the H$_2$ $\nu$=8-6 O(3) line; the weak feature blueward of
  the Pa$\alpha$ line is the He I (4F-3D, 3Fo-3D) line. It is apparent
  that the H$_2$ components do not provide a good fit to the
  Pa$\alpha$ emission, even with the relative fluxes of the components
  allowed to vary freely.}

\label{fig: figs1}

\end{figure}

The molecular hydrogen emission lines show a complex structure of
kinematic components that is not seen in any of the other emission
lines. In the central aperture of the PA105 data, three distinct
components are seen in the H${_2}$ lines --- a central
`narrow' component (FWHM = 160 km s$^{-1}$) and two broader
components (FWHMs = 270 and 370 km s$^{-1}$). Fig. \ref{fig: figs1}
shows these 3 components as fitted to the H$_2$ $\nu$=1-0 S(1) line.
One of the broader components
is blueshifted by (370 $\pm$ 40) km s$^{-1}$ with respect to the systemic
velocity and the other is redshifted by (110  $\pm$ 40) km s$^{-1}$. These
components were derived from the fitting to the H$_2$ $\nu$=1-0 S(1)
line but also describe very well the profile of the H$_2$ $\nu$=1-0
S(2) (see Fig. \ref{fig: figs1}) and, with different intensity
ratios, the H$_2$ $\nu$=1-0 S(4). They also appear to fit the H$_2$
$\nu$=1-0 S(3), but this
is less convincing due to the blend with the [Si VI] line. However, if
the H$_2$ $\nu$=1-0 S(3) line is fit with the H$_2$ $\nu$=1-0 S(1)
components, and the [Si VI] line with the Pa$\alpha$ components, then
the blend can be fitted acceptably. Fig. \ref{fig: figs1} shows that
these H$_2$ components do not provide a good fit to the Pa$\alpha$
emission in the central aperture. These results contradict
those of \citet{wilman00} who found the ortho-hydrogen lines
[S(1) and S(3)] to be significantly broader than the
para-hydrogen lines [S(2) and S(4)] on the nucleus. This discrepancy
may be a result of the single-Gaussian fittings of \citet{wilman00}
being insufficient to fit the actual line profiles.

\begin{figure}

\centerline{\psfig{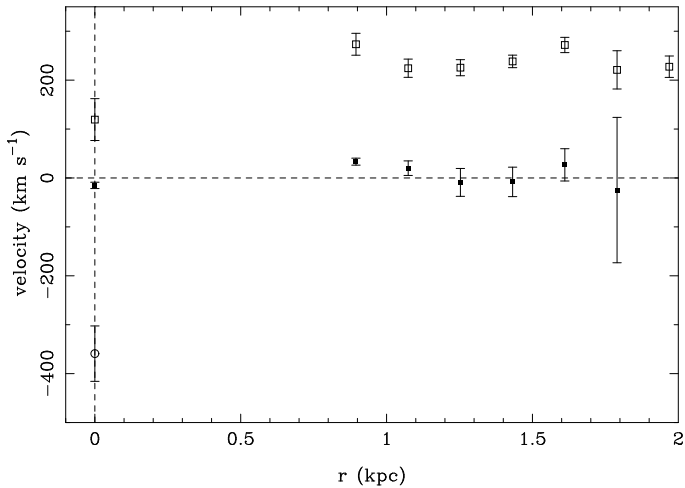}}

\caption{Results of the row-by-row fitting to the H$_2$ $\nu$=1-0 S(1)
line. Two components are seen in the line in the NW aperture and the
variations in the velocities of these components are shown between 0.8
and 2 kpc. The points at $r=0$ show the velocities of the three
components detected in the central nuclear aperture.}

\label{fig: figh2vel}

\end{figure}

As can be seen in Fig. \ref{fig: h2}, the molecular hydrogen shows
a bright, shifted component to the NW of the nucleus. The slit is closely
aligned with the rotation axis of the kpc-scale disc for these observations
and, for this reason, the rotation of this disc would not lead to such
shifted components. The extended structure can be
seen in all the molecular hydrogen lines in the original 2D spectrum,
but the best fit can be made with the bright H$_2$ $\nu$=1-0 S(1) line.
Fitting to this line on a row-by-row basis reveals two
components, one at the systemic velocity and one that is significantly
redshifted. The two components exhibit no
significant velocity variations across the 
range of the fitting. Fig. \ref{fig: figh2vel} shows the velocity
structure of these components over the 0.8-2 kpc region. Fitting on a
row-by-row basis at projected distances $<$0.8 
kpc produces indefinite results without added
constraints and these fits are therefore not plotted. These results
indicate the existence of a molecular hydrogen cloud at a projected
distance of $\sim$1.35
kpc from the nucleus (where the flux of the redshifted component
peaks), redshifted by 243 $\pm$ 7 km s$^{-1}$
in the rest frame. This is consistent with the redshift of 227 $\pm$ 9
km s$^{-1}$ exhibited by one of the on-nucleus H I absorption
components seen by \citet{conway95}. The emission also appears to be
spatially resolved with the flux distribution having a FWHM of
$\sim$500pc (assuming a Gaussian intensity profile).

Overall, the H$_2$ emssion shows kinematic components that are not
seen in other lines. Possible explanations are discussed below.

\section{DISCUSSION}

\subsection{The nuclear emission}

The nuclear kinematics of the Pa$\alpha$ line are undramatic and show
no unusual or unexpected features. Broad (FWHM = 750 km s$^{-1}$) and
narrow (FWHM = 250 km s$^{-1}$) components are detected, each
symmetrically centred at the systemic redshift, with no evidence for
any other components. These line widths are typical of those
seen in the optical data of \citet{taylor03}. The broad component
compares well with the [O III] width integrated over the inner
300 pc (radius) disc as seen in the HST/STIS data of \citet{tadhunter03},
suggesting it represents unresolved rotation and turbulent motions
close to the nucleus.

The narrow component of the H$_2$ lines, although narrower than the
narrow Pa$\alpha$ component (FWHM = 160 km s$^{-1}$), is centred at
the same systemic redshift within the uncertainties. The red
and blueshifted components, however, are not so easily
understood. Their widths, although broader than the central
H$_2$ component, are certainly not individually comparable to the
broad Pa$\alpha$ component. Furthermore, their velocity shifts are
significantly asymmetric about the centre. What is the nature of these
components? Are we seeing 
outflows in the nuclear regions? This is certainly a possibility, but
it is hard to explain how we would then see the redshifted
component. The blueshifted component would obviously correspond to
material flowing out into the cone that is angled towards us and we
would have a relatively clear view of this. The redshifted component,
however, would correspond to an outflow on the far side of the
nucleus and, as it is seen in the nuclear aperture, it would
necessarily lay close to the circum-nuclear torus. It would therefore
be subject to a high degree of extinction, even at near-IR 
wavelengths, and should not be so apparent. The fact that the
redshifted component is the most luminous of the three, and a factor
of 5$\times$ brighter than the blueshifted component, only further compounds
the problem. Another consideration is that, if these components
represent outflows within the ionisation cones, similar components
should be seen in Pa$\alpha$ due to the ionisation of the outflowing
material. Such components are not seen.

It is more likely that these velocity-shifted components originate in
the circumnuclear torus and are linked to its rotation. If, as is
believed, we are viewing Cygnus A at 
close to the opening angle of the NW ionisation cone, then there is
likely to be a relatively clear view of the outer parts of the
circum-nuclear torus. The 
question to ask is why do we see three distinctly separate kinematic
components rather than a continuous broadening of the molecular lines
due to unresolved rotation? It is possible that clumpy obscuring
material partially blocks our view and so we only sample part of the
velocity structure. There is good
evidence that there is patchy dust obscuration within the central
region (\citealt{shaw94}, \citealt*{stockton94}, \citealt{jackson96})
- although on a larger scale - 
so this is by no means an unreasonable explanation.

The H$_2$ $\nu$=1-0 S(3)/H$_2$ $\nu$=1-0 S(1) flux ratio is useful as
a diagnostic indicator of the excitation mechanism of the H$_2$ gas. In the
nuclear aperture this ratio is found to be 1.23 $\pm$ 0.03 (found by
integrating the flux of the two lines over equivalent ranges). X-ray
heating is expected to lead to a ratio of 1.2 (\citealt{lepp83}). Our
result is in good agreement with this and is consistent with the
nuclear H$_2$ gas being excited by emission from the AGN. \citet{tsr99}
previously obtained a value of 1.20 $\pm$ 0.14 for this ratio;
although consistent with our result, this value fails
to rule out shock excitation, expected to result in a ratio of
0.8-1.1 (\citealt{kwan77}; \citealt{hollenbach77}). The more precise
value we have obtained seems to rule out shock excitation as a
plausible mechanism. 

As noted above, the split components visible in the H$_2$ emission are
conspicuously absent in Pa$\alpha$
(see Fig. \ref{fig: figs1}). This is explainable, however, by
noting that the Pa$\alpha$ emission from the torus would originate close to
the ionised inner-edge in a relatively confined region,
whereas the molecular hydrogen would originate in a larger, more
extended region beyond this. Because of the large viewing angle with
respect to the torus axis ($\sim$60\degr), the nearside inner-edge of
the torus lies behind the optically thick outer regions and we would
therefore expect the Pa$\alpha$ to be more highly obscured than the
H$_2$. The Pa$\alpha$ from the 
far inner-edge should be visible however, but this will sample only a
small velocity range and may therefore be detected as part of the
observed broad component.

\subsection{The extended H$_2$ emission}

What is the origin of the redshifted, extra-nuclear H$_2$ component
as seen to the NW of the nucleus (see Fig. \ref{fig: h2} and
Fig. \ref{fig: figh2vel})? There are four possible 
explanations that we will discuss: a gas outflow from the nucleus;
nuclear light scattered off material further out; stellar excitation
of H$_2$ in a star-forming region; and an infalling or
kinematically decoupled gas cloud passing through or close to the
nucleus.

First we shall consider the possibility that this is emission from an
outflowing gas cloud. On the face of it, this is the most apparent and
satisfying explanation: the nuclear winds and radio jets
provide ready-made driving mechanisms for outflows. Such outflows
are an expected by-product of AGN activity and evidence for this in
Cygnus A has already been detected in the [O III] lines
(e.g. \citealt{taylor03};
\citealt{vanbemmel03}). Indirect evidence for this 
phenomenon in Cygnus A is also provided by the hollowed out
cones apparent in near-IR HST images of the core
(\citealt{tadhunter99}). However, there
are a number of problems with this explanation for the shifted H$_2$
component. Firstly, if this cloud
is outflowing within the ionisation cone,
Pa$\alpha$ emission from the photoionised edge of the cloud facing the
nucleus would be observed with similar velocity shifts, but this is
not detected. If this component can be seen in the H$_2$ line it should
certainly be seen in the much brighter
Pa$\alpha$ line. Secondly, since the NW cone is
angled towards us, the far side of the cone lies close to the
plane of the sky ($\lesssim$30\degr): any
outflowing cloud showing a redshift would therefore have to lie
close to the far side of the cone and the necessary de-projected
outflow velocity would be in excess of 450 km s$^{-1}$. It is hard to
envisage how the gas could be accelerated to such high
velocities without being shocked, and shock speeds in excess of
$\sim$50 km s$^{-1}$ would lead to
dissociation of the H$_2$ as well as to Pa$\alpha$ emission
(\citealt{kwan77}; \citealt{draine83}). The combination of these
factors renders an outflow as an unlikely explanation. 

There is also the possibility that the extended emission is nuclear light
scattered off outflowing dust in the cone. In this case the
geometry will always result in a net redshift of the emission unless
the outflowing dust is moving directly towards us along our
line-of-sight (\citealt{vanbemmel03}). This would explain the
redshift, but still
does not explain the lack of an equivalent Pa$\alpha$ component since, to
reflect near-IR light, the scattering material must be optically thick
and therefore relatively grey in its spectral response. This
explanation was proposed in \citet[see section 5.7]{tadhunter03} to
explain the redshifted [O III] emission seen in the NW cone in
HST-STIS optical data; the spectropolarimetry results of
\citet{vanbemmel03} are consistent with this model. However, this [O
  III] emission lies much closer
to the nucleus than the H$_2$ emission ($\sim$200 pc as opposed to
$\sim$1~350 pc) and shows a significantly larger velocity shift
($\sim$450 km s$^{-1}$ as opposed to $\sim$200 km s$^{-1}$). As noted
in \citet{tadhunter03}, there is also a problem with the stability of
any such dust within the ionisation cone and whether it could exist
there without being destroyed, particularly since any gas associated with
the dust must be hot enough to be in an ionised state to avoid being
detected in emission itself.

Alternatively, the H$_2$ emission may arise from the photodissociation
zones associated with H II regions in which stars are
forming; the stars themselves may remain unseen inside the optically
thick clouds. However, near-IR observations of H II galaxies
demonstrate that the Pa$\alpha$ emission from such regions is strong
relative to the H$_2$, with a ratio of Pa$\alpha$/H$_2$~$\nu$=1-0 S(1)
$\sim$5-30 (\citealt{veilleux99}). In contrast, the maximum
ratio allowed by attempted three
component fits to Pa$\alpha$ is Pa$\alpha$/H$_2$~$\nu$=1-0 S(1) $\sim$2
for the redshifted component of H$_2$.
Therefore we conclude that the redshifted H$_2$ component is unlikely
to be excited by a young star forming region.

Finally we consider the case of the H$_2$ emission being associated
with an infalling cloud. Indeed, the possibility
that Cygnus A is undergoing a minor 
merger event (\citealt{canalizo03}) suggests that such components
should be present and natuarally explains the fact that the component
we see is significantly redshifted. Infalling material has already
been seen in the H I 
21 cm absorption components of \citet{conway95} and the redshift of one
of these components is in good agreement with the redshift of the
extended H$_2$ 
component, suggesting a common origin for the two. However,
the problem of
the missing Pa$\alpha$ component remains. If it is falling
through the ionisation cone, the cloud should be photoionised on its
AGN-facing side regardless of the cloud's origin. Alternatively, if it
is outside the ionisation cone, and therefore shielded from the AGN,
then what is stimulating the molecular emission? An intriguing
possibility is that the cloud may be in the foreground outside
the ionisation cone. Here it would be shielded from
the majority of the AGN
emission but would still be illuminated by the more penetrating hard
X-ray radiation. The cloud will appear optically thin to these hard
X-rays resulting in less photoionisation (and hence little
emission from recombination lines such as Pa$\alpha$), but with the
fast electrons released in the rare photoionisation events capable of
thermally exciting large quantities of the molecular gas
(\citealt{lepp83}). This is a plausible explanation for the
discrepancy between the Pa$\alpha$ and H$_2$ emission profiles and one
that rules out a simple outflow, since it is hard to see how the AGN
could be driving outflows beyond the confines of its ionisation cones.

As a test of the infalling cloud premise, the H$_2$ luminosity of the
cloud can be compared to the hard X-ray luminosity of the
AGN. Approximately 0.2\% of the 2-10 keV X-ray flux incident on the
gas cloud will be re-emitted in the H$_2$ $\nu$=1-0 S(1) line
(\citealt{lepp83}), therefore the 2-10keV/H$_2$ $\nu$=1-0 S(1)
luminosity ratio leads directly to an estimate of the covering factor
of the cloud. For our cosmology, the intrinsic 2-10 keV
hard X-ray luminosity of the AGN (deduced from \citealt{young02})
is 1.64 $\times$ 
10$^{44}$ erg s$^{-1}$. If the molecular cloud is outside
the ionisation cone it should see this X-ray emission through a similar
amount of obscuration as we do along our line-of-sight. We therefore
reapply Young et al.'s inferred extinction (column density N$_H$ = 2.0
$\times$ 10$^{23}$ cm$^{-2}$) to derive an obscured luminosity of 1.44
$\times$ 10$^{44}$ erg s$^{-1}$. The integrated luminosity of the
extended H$_2$ $\nu$=1-0 S(1) emission is (4.9 $\pm$ 0.3) $\times$
10$^{39}$ erg s$^{-1}$. Accounting for the 0.2\% efficiency factor,
this implies a cloud covering factor of $\sim$1.7 $\times$
10$^{-2}$. At a projected distance of 1.35 kpc this equates to a cloud
with a circular cross-section $\sim$360 pc in radius. This is large
for a typical {\it single} GMC, but a merger event 
would be expected to involve a large mass of material distributed in
several molecular cloud complexes. The extended
H$_2$ structure appears to be spatially resolved and has an
approximate FWHM of $\sim$500pc wich is in
good agreement with the results of the above calculation.

Certainly, the infalling cloud model seems to be the most
appealing explanation of the four possibilities. The observed redshift
is readily explained and the lack of a matching Pa$\alpha$ component
is not problematic. The missing Pa$\alpha$ alone
is good grounds for eliminating the other explanations and, as
described, there are also other problems associated with them. The
fact that the H$_2$ redshift is in agreement with the redshift of the H I
absorption component is also encouraging. This infalling cloud
scenario also ties in well with the postulated merger event of
\citet{canalizo03}. Unfortunately the S/N of the data is insufficient
for the extended component to be fit to anything other than the H$_2$
$\nu$=1-0 S(1) line, meaning that no information can be obtained about
the excitation of this extended component. However, the motions of the
H$_2$ cloud clearly cannot be
accomodated within the framework of the (equilibrium) kinematics of
the kpc-scale disc associated with the dust lane.

Although the infalling H$_2$ cloud may not be responsible for fuelling
the nuclear activity directly, the process of capture of the cloud, and
subsequent dissipative settling into the kpc-scale disc, will lead to
radial gas motions in the disc and an increased fuelling rate of the
AGN. In such a scenario, the dynamical timescale required for all the
gas associated with the merger event to settle into the disc is likely
to be comparable with the lifetime of the quasar and jet activity.

\section{CONCLUSIONS AND FURTHER WORK}

Our near-IR data have further strengthened the case for the existence of
complex, non-gravitational kinematics in the core of the nearby radio
galaxy Cygnus A. We show several distinct Pa$\alpha$ components, both
on- and off-nucleus, that are consistent with the [O III] components
seen in the optical data of \citet{taylor03}. The molecular hydrogen
emission, however, shows markedly different kinematics that are
inconsistent with both the Pa$\alpha$ and optical data.

We have found good evidence for the existence of an infalling
molecular cloud to the NW of the nucleus and have argued that this
lies outside the main ionisation cone and is being excited by the more
penetrating hard X-ray emission from the AGN. The redshift of this
infalling component 
is in excellent agreement with that of the H I absorption detected by
\citet{conway95}, suggesting the two phenomena are
linked. The fact that the H I absorption is seen on-nucleus, whereas
the H$_2$ emission is seen at a projected radius of $\sim$1.35 kpc,
suggests that the infalling material is likely to have a complex and
extended spatial structure. 

We suggest the possibility that the non-circular motions and infalling
material in the nuclear regions are linked to a possible merger event
(\citealt{canalizo03}) that may have triggered the AGN activity
itself.

However, the problem remains that these long-slit data give
kinematic information only along the radio axis, whereas a minor merger
event could be expected to lead to non-gravitational motion and flows
throughout the cones. Future near-IR integral field data that
map the structure and kinematics of the H$_2$ emission throughout the
cone will reveal if the redhifted emission is associated with the
radio jet itself or is more evenly distributed in the near-nuclear
regions. Near-IR adaptive optics spectroscopy of the nuclear regions
will also be able to clarify the nature of the asymmetrically shifted
H$_2$ components seen in our nuclear aperture.

\section*{ACKNOWLEDGEMENTS}

We acknowledge the facilities and thank the staff at the Keck II
telescope and the Mauna Kea facility. We also thank the referee, Neal
Jackson, for his helpful comments and suggestions. MJB is supported by
a PPARC studentship. We acknowledge the data analysis facilites at
Sheffield provided by the Starlink Project, which is run by CCLRC on
behalf of PPARC.

\bibliographystyle{mn2e}
\bibliography{abbrev,refs}

\begin{thebibliography}{}

\bibitem[\protect\citeauthoryear{Bessell, Castelli \& Plez}{Bessell
  et~al.}{1998}]{bessell98}
Bessell M.~S.,  Castelli F.,    Plez B.,  1998, A\&A, 333, 231

\bibitem[\protect\citeauthoryear{Canalizo, Max, Whysong, Antonucci \&
  Dahm}{Canalizo et~al.}{2003}]{canalizo03}
Canalizo G.,  Max C.,  Whysong D.,  Antonucci R.,    Dahm S.~E.,  2003, ApJ,
  597, 823

\bibitem[\protect\citeauthoryear{Conway \& Blanco}{Conway \&
  Blanco}{1995}]{conway95}
Conway J.~E.,  Blanco P.~R.,  1995, ApJ, 449, L131

\bibitem[\protect\citeauthoryear{Draine, Roberge \& Dalgarno}{Draine
  et~al.}{1983}]{draine83}
Draine B.~T.,  Roberge W.~G.,    Dalgarno A.,  1983, ApJ, 264, 485

\bibitem[\protect\citeauthoryear{Fosbury, Vernet, Villar-Martin, Cohen, Ogle \&
  Tran}{Fosbury et~al.}{1999}]{fosbury99}
Fosbury R. A.~E.,  Vernet J.,  Villar-Martin M.,  Cohen M.~H.,  Ogle P.~M.,
  Tran H.~D.,  1999, in Rottgering H. J.~A.,  Best P.~N.,   Lehnert M.~D.,
  eds, The Most Distant Radio Galaxies, Proceedings of the colloquium,
  Amsterdam, 15-17 October 1997, Royal Netherlands Academy of Arts and Sciences
  p.~311

\bibitem[\protect\citeauthoryear{Hollenbach \& Shull}{Hollenbach \&
  Shull}{1977}]{hollenbach77}
Hollenbach D.~J.,  Shull J.~M.,  1977, ApJ, 216, 419

\bibitem[\protect\citeauthoryear{Jackson, Tadhunter \& Sparks}{Jackson
  et~al.}{1998}]{jackson98}
Jackson N.,  Tadhunter C.,    Sparks W.~B.,  1998, MNRAS, 301, 131

\bibitem[\protect\citeauthoryear{Jackson, Tadhunter, Sparks, Miley \&
  Macchetto}{Jackson et~al.}{1996}]{jackson96}
Jackson N.,  Tadhunter C.,  Sparks W.~B.,  Miley G.~K.,    Macchetto F.,  1996,
  A\&A, 307, L29

\bibitem[\protect\citeauthoryear{Kwan}{Kwan}{1977}]{kwan77}
Kwan J.,  1977, ApJ, 216, 713

\bibitem[\protect\citeauthoryear{Lepp \& McCray}{Lepp \& McCray}{1983}]{lepp83}
Lepp S.,  McCray R.,  1983, ApJ, 269, 560

\bibitem[\protect\citeauthoryear{Ogle, Cohen, Miller, Tran, Fosbury \&
  Goodrich}{Ogle et~al.}{1997}]{ogle97}
Ogle P.~M.,  Cohen M.~H.,  Miller J.~S.,  Tran H.~D.,  Fosbury R. A.~E.,
  Goodrich R.~W.,  1997, ApJ, 482, L37

\bibitem[\protect\citeauthoryear{Shaw \& Tadhunter}{Shaw \&
  Tadhunter}{1994}]{shaw94}
Shaw M.,  Tadhunter C.,  1994, MNRAS, 267, 589

\bibitem[\protect\citeauthoryear{Stockton, Ridgway \& Lilly}{Stockton
  et~al.}{1994}]{stockton94}
Stockton A.,  Ridgway S.~E.,    Lilly S.,  1994, AJ, 108, 414

\bibitem[\protect\citeauthoryear{Tadhunter, Marconi, Axon, Wills, Robinson \&
  Jackson}{Tadhunter et~al.}{2003}]{tadhunter03}
Tadhunter C.,  Marconi A.,  Axon D.,  Wills K.,  Robinson T.~G.,    Jackson N.,
   2003, MNRAS, 342, 861

\bibitem[\protect\citeauthoryear{Tadhunter}{Tadhunter}{1991}]{tadhunter91}
Tadhunter C.~N.,  1991, MNRAS, 251, 46P

\bibitem[\protect\citeauthoryear{Tadhunter, Packham, Axon, Jackson, Hough,
  Robinson, Young \& Sparks}{Tadhunter et~al.}{1999}]{tadhunter99}
Tadhunter C.~N.,  Packham C.,  Axon D.~J.,  Jackson N.~J.,  Hough J.~H.,
  Robinson A.,  Young S.,    Sparks W.,  1999, ApJ, 512, L91

\bibitem[\protect\citeauthoryear{Taylor, Tadhunter \& Robinson}{Taylor
  et~al.}{2003}]{taylor03}
Taylor M.~D.,  Tadhunter C.~N.,    Robinson T.~G.,  2003, MNRAS, 342, 995

\bibitem[\protect\citeauthoryear{Thornton, Stockton \& Ridgway}{Thornton
  et~al.}{1999}]{tsr99}
Thornton R.~J.,  Stockton A.,    Ridgway S.,  1999, AJ, 118, 1461

\bibitem[\protect\citeauthoryear{Ueno, Koyama, Nishida, Yamauchi \& Ward}{Ueno
  et~al.}{1994}]{ueno94}
Ueno S.,  Koyama K.,  Nishida M.,  Yamauchi S.,    Ward M.~J.,  1994, ApJ, 431,
  L1

\bibitem[\protect\citeauthoryear{van Bemmel, Vernet, Fosbury \& Lamers}{van
  Bemmel et~al.}{2003}]{vanbemmel03}
van Bemmel I.~M.,  Vernet J.,  Fosbury R. A.~E.,    Lamers H. J. G. L.~M.,
  2003, MNRAS, 345, L13

\bibitem[\protect\citeauthoryear{Veilleux, Sanders \& Kim}{Veilleux
  et~al.}{1999}]{veilleux99}
Veilleux S.,  Sanders D.~B.,    Kim D.-C.,  1999, ApJ, 522, 139

\bibitem[\protect\citeauthoryear{Ward, Blanco, Wilson \& Nishida}{Ward
  et~al.}{1991}]{ward91}
Ward M.~J.,  Blanco P.~R.,  Wilson A.~S.,    Nishida M.,  1991, AJ, 382, 115

\bibitem[\protect\citeauthoryear{Wilman, Edge, Johnstone, Crawford \&
  Fabian}{Wilman et~al.}{2000}]{wilman00}
Wilman R.~J.,  Edge A.~E.,  Johnstone R.~M.,  Crawford C.~S.,    Fabian A.~C.,
  2000, MNRAS, 318, 1232

\bibitem[\protect\citeauthoryear{Young, Wilson, Terashima, Arnaud \&
  Smith}{Young et~al.}{2002}]{young02}
Young A.~J.,  Wilson A.~S.,  Terashima Y.,  Arnaud K.~A.,    Smith D.~A.,
  2002, ApJ, 564, 176

\end{thebibliography}

\end{document}